\documentclass[12pt]{article}
\date{}
\title{\textbf{\Large{The Klein-Gordon Equation for the
Coulomb Potential in Non-commutative Space}}}
\author{{\small{Amin Rezaei Akbarieh and Hossein Motavalli}}\\
\small{Faculty of Physics, University of Tabriz, Tabriz 56554,
Iran.}\\
\small{Motavalli@Tabrizu.ac.ir}}
\begin{document}
\maketitle

\section{Abstract}
{In this paper the stationary Klein-Gordon equation is considered
for the Coulomb potential in non-commutative space. The energy shift
due to non-commutativeity is obtained via the perturbation theory.
Furthermore, we show that the degeneracy of the initial spectral
line is broken in transition from commutative space to
non-commutative space.}\\

Keywords: Klein-Gordon equation; Coulomb potential; Non-commutative
\\PACS Nos.: 03.65.-w; 03.65.Ge; 03.65.Ta.
\section{Introduction}
Recently, there has been an increased interest in the study of the
non-commutative field theory [1-2]. The most important motivation
for studying these theories, comes mainly from the works that
establish a connection between non-commutative geometry and string
theory [3]. The investigation of these theories gives us the
opportunity to understand interesting phenomena, such as
non-locality and IR/UV mixing [4], new physics at very short
distances [1-2], and possible implications of Lorentz violation
[5-6]. Among these theories, the quantum mechanics is one of the
simplest theories [7-8]. It is well-known that solutions of the
relativistic wave equation play an essential role in the
relativistic quantum mechanics for some physical potentials of
interest [9-13]. Recently, there has been an increasing interest in
finding exact solutions of the Klein-Gordon (KG) equation [14-18].
In the past few years, exact solutions and energy eigenvalues of
this equation have been presented for Scarf [19], Rosen-Morse type
[20], Hulthen [21], Wood-saxon [22, 23], Posch-Teller [24],
five-parameter exponential [25, 26], generalized symmetrical
double-well [27], ring-shape harmonic oscillator [28], and pseudo
harmonic oscillator [29] potentials, etc. In the above cited papers
the scalar and vector potentials are almost taken to be equal in the
relativistic framework. However, there is almost no explicit
expression for the energy eigenvalues. Within the framework of
non-commutativity, situation is more complicated and most models
cannot be solved exactly. Accordingly, most of the available results
are based upon perturbation theory [30-31]. This implies that a
simple physical system in the commutative space may be changed into
a complex theory within non-commutative framework.

Inclusion of non-commutativity into the quantum field theory can be
achieved in two different ways: via Moyal product on the space of
ordinary functions, or redefining the field theory on a coordinate
operator space which is intrinsically non-commutative [32-33]. The
equivalence between the two approaches has been described in
references [34-35]. In the usual method, we introduce
non-commutativity by means of non-commutative coordinates of
position and momentum $(x, p)$ satisfying the following commutation
relations
\begin{eqnarray}
[x_i\;,\;x_j]=i{\theta}_{ij}\;\;,\;\;[x_i \;, \; p_j]=i
{\delta}_{ij} \;\; , \;\; [p_i \; , \; p_j]=0, \;\; i,j=1,2,3
\end{eqnarray}
where ${\theta}_{ij}={\epsilon}_{ij}\theta$, in which
${\epsilon}_{ij}$ is Levichevita symbol and $\theta$ is a
parameter that measures the non-commutativity of coordinates. In
the non-commutative space the ordinary product is replaced by
Moyal product
\begin{eqnarray}
f(x)\star
g(x)=exp\{\frac{i}{2}{\theta}^{jk}{\partial_j}^{(1)}{\partial_k}^{(2)}\}
f(x_1)g(x_2)|_{x_1=x_2=x} \nonumber
\end{eqnarray}
where $f(x)$ and $g(x)$ are two arbitrary differentiable
functions.

\section{The Non-commutative Klein-Gordon Equation}
\label{sec:1}

In this section we consider the three dimensional Klein-Gordon equation for a
long-range $1/r$ interaction in the non-commutative space. For time independent
potentials, the KG equation for a particle of rest mass $M$ can be written as
($\hbar$=c=1)
\begin{eqnarray}
\{ {\nabla}^2+[V(r)-E]^2-[S(r)+M]^2\}\psi(r)=0
\end{eqnarray}
in commutative space, where $E$ is the relativistic energy, $V(r)$
and $S(r)$ denote vector and scalar potentials, respectively.
Recently, interest for considering of this equation with equal
scalar and vector potentials has been increased [19-20]. Under
assumption $V(r)=S(r)$, Eq. (2) takes the form
\begin{eqnarray}
\{ {\nabla}^2+(E^2-M^2)-2(E+M)V(r)\}\psi(r)=0.
\end{eqnarray}
By using the common separation of variables in the spherical polar coordinate
$\psi(r)=Y(\theta,\phi)R(r)/r$, the radial part of this equation reads
\begin{eqnarray}
\{\frac{d^2}{dr^2}-[E_{eff}+V_{eff}(r)]\}R(r)=0
\end{eqnarray}
 where
\begin{eqnarray}
V_{eff}(r)=2(M+E)V(r)+\ell(\ell+1)/r^2, \;\;\;\;\;E_{eff}=(M^2-E^2).
\end{eqnarray}
Now to consider this equation in the non-commutative space, let us introduce the
non-commuting coordinates in terms of the commuting coordinates and
their momenta
\begin{eqnarray}
\left\{
  \begin{array}{ll}
    \hat{x}_i=x_i+\frac{1}{2}\theta_{ij}p_j, \\
    \hat{p}_i=p_i.
  \end{array}
\right.
\end{eqnarray}
Under these transformations a radial form potential takes the form
\begin{eqnarray}
V(\hat{r}) \nonumber &=&V(|\vec{r}-\frac{\vec{p}}{2}|)\\ \nonumber
&=&V(\sqrt{(x_i-\frac{1}{2}\theta_{ij}p_j)(x_i-\frac{1}{2}\theta_{ij}p_j)}\;\;)\\
\nonumber &=&V(r)+\frac{1}{2}(\vec{\theta}\times \vec{p})\cdot
\vec{\nabla}V(r)+O(\theta^2)\\ \nonumber
&=&V(r)-\frac{\vec{\theta} \cdot \vec{L}}{2 r}\frac{\partial V}{\partial r}+O(\theta^2)\\
&\simeq & V(r)-\frac{\vec{\theta} \cdot \vec{L}}{2 r}\frac{\partial
V}{\partial r}
\end{eqnarray}
up to the first order of $\theta$, where$\;\;\;r=\sqrt{x_ix_i}$ and
$\vec{L}=\vec{r}\times\vec{p}$ is the angular momentum operator. \\
By replacement of the ordinary product with Moyal, Eq.(6) takes
the following form
\begin{eqnarray}
\{\frac{d^2}{dr^2}-[E_{eff}+V_{eff}(r)]\}\star R_{n\ell}(r)=0 \nonumber
\end{eqnarray}
in the non-commutative space, or equivalently
\begin{eqnarray}
\{\frac{d^2}{dr^2}-[E_{eff}+V_{eff}(|\vec{r}-\frac{1}{2}{\vec{p}}|)]\}R_{n\ell}(r)=0.
\end{eqnarray}
Comparing Eq. (6) with Eq. (8) indicates that under the Moyal
product the only modification in the radial part of the KG equation
appears in the effective potential term. By substituting Coulomb
potential $V(r)=-\frac{Ze^2}{r}$  into relation (5) and using
effective potential (7) the last equation can be rewritten as
{\footnotesize\begin{eqnarray}
\{\frac{d^2}{dr^2}-\frac{\ell(\ell+1)}{r^2}
+2(E+M)\frac{Ze^2}{r}-E_{eff}-\frac{(\vec{\theta}\cdot \vec{L})}{2
r}[\frac{2\ell(\ell+1)}{r^3} -2(E+M)\frac{Ze^2}{r^2}] \} R(r)=0.
\end{eqnarray}}
By introducing dimensionless new variable $\rho=2r\sqrt{E_{eff}}$,
Eq. (9) is transformed into the following form
{\footnotesize\begin{eqnarray}
\{\frac{d^2}{d\rho^2}-\frac{\ell(\ell+1)}{\rho^2}+
\frac{\varsigma}{\rho}-\frac{1}{4}-(\vec{\theta}\cdot
\vec{L})[\frac{4\ell(\ell+1)E_{eff}}{\rho^4}
-2(1+E/M)\sqrt{E_{eff}}\frac{Z\alpha}{\rho^3}]\}R(\rho)=0
\end{eqnarray}}
where $\;\varsigma=\frac{Z\alpha}{M}\sqrt{1+\frac{2E}{M-E}}$.

\section{The Solution}
\label{sec:1} The last equation has not yet been solved exactly in
the presence of the last two terms, whereas in their absence, its
exact solution is available [36]. To obtain the solution, we choose $\theta=0$, and get
\begin{eqnarray}
\{\frac{d^2}{d\rho^2}-\frac{\ell(\ell+1)}{\rho^2}+\frac{\varsigma}{\rho}-\frac{1}{4}\}R^{(0)}(\rho)=0.
\end{eqnarray}
This is a second order differential equation and can be easily
solved via Nikiforov-Uvarov (NU) mathematical method. In this method
a second order linear differential equation is reduced to a
generalized equation of hyper-geometric type whose exact solutions
are expressed in terms of special orthogonal functions [37], as well
as corresponding eigenvalues are obtained. To apply this method for
Eq. (11), we compare this equation with the generalized
hyper-geometric type equation
\begin{eqnarray}
\{\frac{d^2}{d\rho^2}+\frac{\tilde{\tau}(\rho)}{\sigma(\rho)}\frac{d}{d\rho}+\frac{\tilde{\sigma}(\rho)}{\sigma^{2}(\rho)}\}R^{(0)}(\rho)=0
\end{eqnarray}
and get
\begin{eqnarray}
\tilde{\tau}(\rho)=0,   \;\;\; \sigma(\rho)=2\rho,\;\;\;
\tilde{\sigma}(\rho)=-4\ell(\ell+1)-\rho^2+4\varsigma\rho.
\end{eqnarray}
Using these functions it is straightforward to show that the exact
solution of Eq. (11) is [19]
\begin{eqnarray}
R^{(0)}(\rho)=N\rho^{\ell+1}\frac{(n-\ell-1)!}{(n+\ell)!}(2\ell+1)!L^{2\ell+1}_{n-\ell-1}(\rho)e^{-\frac{\rho}{2}},
\;\;\;\;\;\;n=0,\;1,\;2,...
\end{eqnarray}
where $L^{2\ell+1}_{n-\ell-1}(\rho)$ denotes the generalized Laguerre polynomials and
$N$ is normalization constant
\begin{eqnarray}
N=\sqrt{\frac{(n+\ell)!}{2|E^{(0)}|n(n-\ell-1)!}}\frac{1}{(2\ell+1)!}
\end{eqnarray}
in which $E^{(0)}$ is the energy eigenvalues and is given by
\begin{eqnarray}
E^{(0)}=\{\frac{(Z\alpha)^2-(n-\ell)^2M^2}{(Z\alpha)^2+(n-\ell)^2M^2}\}M, \;\;\;\;\;\;n=0,\;1,\;2,....
\end{eqnarray}
Now, to obtain the modifacation of energy levels as a result of the
last two terms in Eq. (10) due to the non-commutativity, we use perturbation theory.
For simplicity, first of all we take ${\theta}_3={\theta}$ and assume that the
other ${\theta}$-components are zero (by rotation or redefinition of coordinates),
such that $\vec{\theta} \cdot\vec{L}=L_z {\theta}$. In addition, we use
\begin{eqnarray}
<nlm|L_z|nlm'>=m{\delta}_{mm'} ,\;\;\;\;\;\;\;\;\;\;\;\;\;\;-l\leq
m\leq l\nonumber
\end{eqnarray}
and also the fact that in the first order perturbation theory the
expectation value of $\rho^{-3}$ and $\rho^{-4}$ with respect to the
exact solution of Eq. (11), are given by [38]
\begin{eqnarray}\nonumber
<n|{\rho^{-3}}|n>&=&\int \{R^{(0)}(\rho)\}^2{\rho}^{-1}d\rho=\frac{1}{2|E^0|}\frac{1}{\ell(2\ell+1)(2\ell+2)} \\\nonumber
<n|{\rho^{-4}}|n>&=&\int \{R^{(0)}(\rho)\}^2{\rho}^{-2}d\rho=
\frac{1}{n|E^0|}\frac{\Gamma(2\ell-1)}{\Gamma(2\ell+4)}[3n^2-\ell(\ell+1)].
\end{eqnarray}
Putting these results together, one gets
{\footnotesize\begin{eqnarray}\nonumber \Delta
E_{NC}=\frac{m\theta}{4(2\ell+1)|E^0|} \{\frac{(3n^2-\ell(\ell+1))
}{n(2\ell-1)(2\ell+3)}-\frac{2(n-\ell)^2
Z\alpha}{\ell(\ell+1)[(n-\ell)^2+(Z\alpha/M)^2]}
\},\;\;\;\;\;n=0,1,2,....
\end{eqnarray}}
This is energy shift due to the additional  last two terms of Eq.
(10). The appearance of the magnetic quantum number $m$ in this
expression explicitly indicates the splitting of states with the
same orbital angular momentum into the corresponding components. In
fact each level $\ell$ splits into $2\ell+1$ sublevels and subsequently breaks
the degeneracy of the initial spectral line. The lifting of
degeneracy is due to the emergence of a magnetic field associated
with the non-commutative space in transition from commutative space
into non-commutative space. This behavior is similar to the Zeeman
effect. In addition, it is worth noting that the correction terms
containing $\vec{\theta}\cdot \vec{L}$ are very similar to that of
the spin orbit coupling, in which the non-commutative parameter
$\vec{\theta}$ plays the role of the spin.

\section{Conclusion}
In this paper, we have investigated the Klein-Gordon equation for
the Coulomb potential in the non-commutative space. The energy
shift, due to the non-commutativity, is obtained via first order
perturbation theory. It is explicitly shown that the degeneracy of
the initial spectral line is broken in transition from commutative
space into non-commutative space by splitting states into the
corresponding components. This behavior is similar to the Zeeman
effect in which a magnetic field is applied to the system. In this space
the non-commutative parameter $\vec{\theta}$ plays the role of the
spin.

\end{document}